\documentclass[times,doublespace]{ettauth}

\usepackage{moreverb}

\usepackage[colorlinks,bookmarksopen,bookmarksnumbered,citecolor=red,urlcolor=red]{hyperref}

\newcommand\BibTeX{{\rmfamily B\kern-.05em \textsc{i\kern-.025em b}\kern-.08em
T\kern-.1667em\lower.7ex\hbox{E}\kern-.125emX}}

\usepackage{amssymb}
\usepackage{amsmath}
\usepackage{algorithmic}
\usepackage{algorithm}
\usepackage{psfrag}
\usepackage{graphicx}
\usepackage{cleveref}
\usepackage{cite}

\DeclareMathAlphabet{\mathbit}{OML}{cmr}{bx}{it}
\newcommand{\B}[1]{\mathbit{#1}}
\DeclareMathOperator{\Exp}{\mathbb{E}}
\DeclareMathOperator{\Transpose}{T}
\DeclareMathOperator{\Hermitian}{H}
\DeclareMathOperator{\trace}{tr}
\DeclareMathOperator{\diag}{diag}

\DeclareMathOperator{\Fr}{F}

\newcommand{\Tr}{{\Transpose}}
\newcommand{\He}{{\Hermitian}}
\newcommand{\bs}{\boldsymbol}
\newcommand{\MSE}{\text{MSE}}
\newcommand{\MMSE}{\text{MMSE}}
\newcommand{\gt}{\tilde{\B{g}}}
\newcommand{\imj}{j}
\DeclareMathOperator{\eul}{e}

\begin{document}

\runningheads{J.~P.~Gonz\'alez-Coma et al.}{QoS Optimization in the Multiple Stream MIMO Broadcast Channel}

\articletype{RESEARCH ARTICLE}

\title{QoS Constrained Power Minimization in the Multiple Stream MIMO Broadcast Channel}

\author{Jos\'e~P.~Gonz\'alez-Coma\affil{1}\corrauth, Michael~Joham\affil{2}, Paula~M.~Castro\affil{1} and Luis~Castedo\affil{1}}

\address{\affilnum{1}Universidade da Coru\~na, Facultad de Inform\'atica, Campus Elvi\~na s/n, A Coru\~na,  15071, Spain\\ \affilnum{2}Technische Universit\"at M\"unchen, Deparment of Electrical and Computer Engineering, Theresienstrasse 90 M\"unchen, 80333, Germany}

\corraddr{Universidade da Coru\~na, Facultad de Inform\'atica, Campus Elvi\~na s/n, A Coru\~na,  15071, Spain\\E-mail: jose.gcoma@udc.es}

\begin{abstract}
This work addresses the design of optimal linear transmit filters for the \emph{Multiple Input-Multiple Output} (MIMO) \emph{Broadcast Channel} (BC) when several spatial streams are allocated to each user. We also consider that the \emph{Channel State Information} (CSI) is perfect at the receivers but is only partial at the transmitter. A statistical model for the partial CSI is assumed and exploited for the filter design. Similarly to the single-stream per user case, the problem is solved via \emph{Mean Square Error} (MSE) dualities and interference functions. However, including more streams per user involves an additional complexity level since we must determine how to distribute the per-user rates among the streams. Such problem is solved using a projected gradient algorithm.
\end{abstract}


\maketitle


\section{Introduction}
This work focuses on power minimization in the \emph{Multiple Input-Multiple Output} (MIMO) \emph{Broadcast Channel} (BC) when several streams are allocated to each user. Considering more than a single stream per user takes advantage of the spatial multiplexing gain of MIMO systems to increase the communication rate. Furthermore, multiple data streams fit current scenarios where users connect more than one devices, or different and simultaneous data streams are required for a certain number of applications running at one device. 

Our goal is to minimize the total amount of power needed to fulfill certain per-user \emph{Quality-of-Service} (QoS) restrictions, also taking into account the flexibility of distributing the rate constraints between the different per-user streams. By imposing these restrictions, we ensure that all users achieve certain level of data rate. This is in contrast to sum-rate maximizations \cite{NeGhSl}, or max-min formulations \cite{ShScBo08} where users with poor channels obtain low data rates. Note that the \emph{Base Station} (BS) has usually more degrees of freedom than the individual receivers. Therefore, it is appropriate to mitigate the interference between users by precoding at the transmitter.  

Regarding \emph{Channel State Information} (CSI), we consider that perfect knowledge is available at the receivers but only partial at the BS. This makes a difference with respect to previous works \cite{KobCai07,CodTolJunLat,WeLa,ViJiGo,ShScBo07,HeJoUt} where perfect knowledge of the CSI at both ends of the BC is assumed. Moreover, not purely stochastic nor bounded error models are assumed as in \cite{ShDa07,MuKiBo07,ZhWoOt09,NeGhSl,RaBoZh}, but a statistical knowledge of the available CSI at the BS by means of a \emph{probability density function} (pdf).

In this work, contrary to \cite{GoJoCaCa13ICASSP,JoCl14}  where a \emph{Multiple Input-Single Output} (MISO) BC with only one stream per user is considered, we take advantage of the MIMO BC spatial multiplexing capabilities to send more than one spatial streams to each user. Thus, the dimensions of both the transmit and receive filters have to be adapted accordingly. Moreover, considering more than one per-user streams has a high impact on the problem formulation. Indeed, the \emph{Minimum Mean Square Error} MMSE-based lower bound employed in previous works, e.g. \cite{GoJoCaCa13ICASSP}, is not tight in the considered scenario. An additional complexity level arises  since the designer has to decide between different per-stream target rates fulfilling the same per-user target rates \cite{GoJoCaCa14EUSIPCO,GoJoCaCa14SAM}. Contrary to \cite{GoJoCaCa14EUSIPCO,GoJoCaCa14SAM}, we consider the possibility of switching on and off part of the per-user streams for the problem solution to lie in the feasibility region. Finally, a discussion on the sum-MMSE region provides insight about the relationship with the rate restrictions.

The paper is organized as follows. Section \ref{sec:sysmodel} describes the MIMO BC system model when several data streams are allocated to each of the users. Section \ref{sec:MSprobForm} addresses some transformations of the original problem. The proposed solution is presented in Section \ref{sec:projgrad}. Finally, the results of simulation experiments are given in Section \ref{sec:sim} and the conclusions in Section \ref{sec:conc}.  

The following notation is employed. Matrices and column vectors are written using upper an lower boldface characters, respectively. By $[\B{X}]_{j,k}$, we denote the element in row $j$ and column $k$ of the matrix $\B{X}$; $\diag(x_i)$ represents a diagonal matrix whose $i$th diagonal element is $x_i$; $\mathbf{I}_N$ stands for the $N\times N$ identity matrix, and $\B{e}_i$ represents the canonical vector. The superscripts $(\cdot)^*$, $(\cdot)^\Tr$, and $(\cdot)^\He$ denote the complex conjugate, transpose, and Hermitian. $\Re\{\cdot\}$ represents the real part operator.  Finally, $\Exp[\cdot]$ stands for statistical expectation, $\trace(\cdot)$ and $\det(\cdot)$ denote the trace and determinant operators, and $|\cdot|$, \linebreak$\|\cdot\|_2$, $\|\cdot\|_{\Fr}$ stand for the absolute value, the Euclidean norm, and the Frobenius norm, respectively.
\section {System Model}
\label{sec:sysmodel}
\begin{figure}[!t]
  \centering
	\psfrag{s1}{$\B{s}_1$} 
	\psfrag{sK}{$\B{s}_K$} 
	\psfrag{p1}{$\B{P}_1$}
	\psfrag{pK}{$\B{P}_K$}
	\psfrag{h1}{$\B{H}_1^{\He}$}
	\psfrag{hK}{$\B{H}_K^{\He}$}
	\psfrag{n1}{$\B{\eta}_1$}
	\psfrag{nK}{$\B{\eta}_K$}
	\psfrag{y}{}
	\psfrag{x1}{}
	\psfrag{xK}{}
	\psfrag{sb1}{$\hat{\B{s}}_1$}
	\psfrag{sbK}{$\hat{\B{s}}_K$}
	\psfrag{b1}{$\B{F}_1^{\He}$}
	\psfrag{bK}{$\B{F}_K^{\He}$}
	\includegraphics[width=\linewidth]{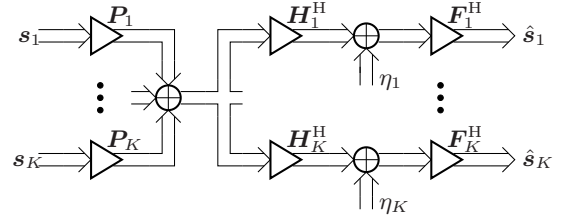}
\caption{Multiple Stream MIMO BC System Model.}
\label{fig:msmimobc}
\end{figure}
\vspace{-0.2cm}	
Fig. \ref{fig:msmimobc} shows the block diagram of a multiple stream MIMO BC. $K$ users, with $R$ antennas each, receive the information sent from a BS with $N$ antennas. The data symbols are represented by the vectors $\B{s}_k\in \mathbb{C}^{d_k}$ comprising the $d_k$ data streams transmitted to the $k$th user, $k\in\{1,\ldots,K\}$. Such data vectors are considered to be zero-mean Gaussian with a covariance matrix $\Exp[\B{s}_k\B{s}_k^{\He}]=\mathbf{I}_{d_k}$, i.e. $\B{s}_k\sim\mathcal{N}_\mathbb{C}(\mathbf{0},\mathbf{I}_{d_k})$, and  independent among users, i.e. $\Exp[\B{s}_k\B{s}_l^{\He}]=\mathbf{0}$ for $l \ne k$. Prior to be transmitted, the data vectors are precoded with $\B{P}_k\in \mathbb{C}^{N\times d_k}$ to produce the signal that propagates over the MIMO channel $\B{H}_k\in \mathbb{C}^{N\times R}$. The signal received by each user is then filtered with a linear receive filter $\B{F}_k\in\mathbb{C}^{R\times d_k}$ to produce the data estimates 
\begin{equation}
\hat{\B{s}}_k=\B{F}_k^\He\B{H}_k^{\He}
\sum\nolimits_{i=1}^K\B{P}_i\B{s}_i+\B{F}_k^\He\bs{\eta}_k,
\label{eq:MSMIMOestimates}
\end{equation}
where $\bs{\eta}_k \sim \mathcal{N}_{\mathbb{C}}(\mathbf{0},\B{C}_{\bs{\eta}_k})$ is the $k$th user's zero-mean additive Gaussian noise with covariance matrix $\B{C}_{\bs{\eta}_k}$. According to such system model, the $k$-th user data rate is given by
\begin{equation}
R_k=\log_2\det\left(\mathbf{I}_R+\B{H}_{k}^\He\B{P}_{k}\B{P}_{k}^\He\B{H}_{k}\B{X}_k^{-1}\right),
\label{eq:rate}
\end{equation} 
where $\B{X}_k=\B{C}_{\bs{\eta}_k}+\B{H}_{k}^\He\sum_{i\neq k}\B{P}_{i}\B{P}_{i}^\He\B{H}_{k}$ represents the interference from the other users and the noise. The total transmit power is $P_T=\sum_{k=1}^{K}||\B{P}_k||_{\Fr}^2$.

Expression \eqref{eq:rate} is appropriate for perfect CSI at the BS. However, we consider that the CSI at the BS, $v$, is partial and available through the conditional pdfs $f_{\B{H}_k\vert v}\left(\B{H}_k\vert v\right)$. No additional assumptions regarding $f_{\B{H}_k\vert v}\left(\B{H}_k\vert v\right)$ are made. Thus, the QoS metric is given by the $k$th user conditional average rate
\begin{equation}
\Exp[R_k|\,v]=\Exp\left[\log_2\det\left(\mathbf{I}_R+\B{H}_{k}^\He\B{P}_{k}\B{P}_{k}^\He\B{H}_{k}\B{X}_k^{-1}\right)|\,v\right]
\label{eq:MUMIMOrate}
\end{equation} 
be larger than a given value $\rho_k$. This leads to the following optimization problem
\begin{align}
\min_{\left\{\B{P}_k(v)\right\}_{k=1}^K} P_T&=\sum_{k=1}^K
\left\|\B{P}_k(v)\right\|_\text{F}^2\nonumber\\
&\quad\text{s.t.}\quad \Exp\left[R_k|\,v\right]\geq\rho_k\,\forall k,
\label{eq:BCform}
\end{align}
where we remarked the dependency of $\B{P}_k$ on $v$. In the ensuing section we explain how to solve the optimization problem \eqref{eq:BCform} by exploiting the relationship between rate and MMSE to rewrite the constraints in a more manageable way. Recall that the approximations employed in \cite{GoJoCaCa13ICASSP,JoCl14,GoJoCaCA13_2} are not applicable here.

\section{Problem Formulation}
\label{sec:MSprobForm} 
Let us introduce the multiple-stream BC MSE as follows
\begin{align}
\text{MSE}_k^\text{BC}&=\left\|\B{s}_k-\hat{\B{s}}_k\right\|^2_2=\trace\left(\mathbf{I}_{d_k}-2\Re\left\{\B{F}_k^\He\B{H}_k^\He\B{P}_k\right\}\right)\nonumber\\
&+\sum_{i=1}^{K}\left\|\B{F}_k^\He\B{H}_k^\He\B{P}_i\right\|^2_{\Fr}+\trace\left(\B{F}_k^\He\B{C}_{\bs{\eta}_k}\B{F}_k\right).
\label{eq:MUMIMOMSEBC}
\end{align}
Since the CSI is imperfect at the BS, the appropriate MSE measure is the conditional average $\Exp[\MSE|\, v] = \overline{\MSE}^{\text{BC}}_{k}(v)$. This is in accordance with the rate in \eqref{eq:MUMIMOrate}.

Recall, however, that CSI is perfect at the receiver-side and hence we can determine the MMSE receive filters for given precoders $\B{P}_k(v)$, i.e.,
\begin{equation}
  \B{F}_k^{\text{MMSE}}=\left(\B{H}_k^\He\B{P}_k(v)\B{P}_k^\He(v)\B{H}_k+\B{X}_k\right)^{-1}
		\B{H}_k^\He\B{P}_k(v),
		\label{eq:MSMIMOFmmse}
\end{equation}
with $\B{X}_k=\B{H}_k^\He\sum_{i\neq k}\B{P}_i(v)\B{P}_i^\He(v)\B{H}_k+\B{C}_{\bs{\eta}_k}$.
Plugging \eqref{eq:MSMIMOFmmse} into \eqref{eq:MUMIMOMSEBC} yields the following expression for the $k$-th user average minimum MSE
\begin{align}
&\overline{\MMSE}\mbox{}^{\text{BC}}_{k}(v)=\Exp\left[\trace\left(\B{\Sigma}_k(v)\right)|\,v\right]\label{eq:MMSESigma}\\
&=\Exp\left[\trace\left(\mathbf{I}_{d_k}+\B{P}_k^{\He}(v)\B{H}_k\B{X}_k^{-1}\B{H}_k^{\He}\B{P}_k(v)\right)^{-1}|\,v\right].\nonumber
\end{align} 

Observe that applying Sylvester's determinant identity to \eqref{eq:MUMIMOrate}, the average rate is a function of $\B{\Sigma}_k(v)$, as follows  
\begin{equation}
\Exp\left[R_k|\,v\right]=\Exp\left[\log_2\det\left(\B{\Sigma}_k^{-1}(v)\right)\,|\,v\right].
\label{eq:RateSigma}
\end{equation}

Notice the positive semidefiniteness of the matrix product $\B{P}_k^{\He}(v)\B{H}_k\B{X}_k^{-1}\B{H}_k^{\He}\B{P}_k(v)$ in \eqref{eq:MMSESigma}. Therefore, $\B{\Sigma}_k(v)$ in \eqref{eq:MMSESigma} and \eqref{eq:RateSigma}, and also $\Exp\left[\B{\Sigma}_k(v)|\,v\right]$, are positive semidefinite matrices. Thus, we compute the eigenvalue decomposition $\Exp[\B{\Sigma}_k(v)|\,v]=\B{U}_k\B{\Lambda}_k\B{U}_k^{\He}$, with the unitary matrix $\B{U}_k$ and the diagonal matrix $\B{\Lambda}_k=\diag(\lambda_{k,1},\ldots,\lambda_{k,d_k})$, where $\lambda_{k,i}\geq0,\,\forall k,i$ are the eigenvalues.

The columns of $\B{U}_k$ form a basis that enables to introduce the spatial decorrelation precoders $\B{P}'_k(v)=\B{P}_k(v)\B{U}_k$. Such precoders remove the off-diagonal elements of $\Exp[\B{\Sigma}_k(v)|\,v]$ for all $k$, without changing the total transmit power $\sum_{k=1}^{K}\left\|\B{P}_k^\prime(v)\right\|^2_{\Fr}=\sum_{k=1}^{K}\left\|\B{P}_k(v)\right\|^2_{\Fr}$, nor the expressions of the average rate \eqref{eq:MUMIMOrate} and the average MMSE \eqref{eq:MMSESigma}. We henceforth consider that the spatial decorrelation precoders $\B{P}'_k$ are employed. Thus, the per-user average MMSE in the BC is
\begin{equation}
\overline{\MMSE}\mbox{}^{\text{BC}}_k(v)=\trace\left(\Exp\left[\B{\Sigma}_k(v)|v\right]\right)=\sum_{i=1}^{d_k} \lambda_{k,i}.
\label{eq:diagMMSE}
\end{equation}
Notice that $\lambda_{k,i}$ can be interpreted as the $k$-th user $i$-th stream average MMSE, i.e. $\lambda_{k,i} = \overline{\text{MMSE}}^{\text{BC}}_{k,i}(v)$, and the average MMSE in \eqref{eq:diagMMSE} corresponds to the sum of such individual average MMSEs, i.e. $\overline{\text{MMSE}}^{\text{BC}}_{k}(v)=\sum_{i=1}^{d_k} \overline{\text{MMSE}}^{\text{BC}}_{k,i}(v)$.

The function $f(\B{A})=-\log(\det(\B{A}))$, with $\B{A}$ being positive semidefinite, is a convex function. Hence, applying Jensen's inequality to \eqref{eq:RateSigma} gives 
\begin{align}
  \Exp[R_k|\,v]&\ge
	-\log_2\det\left(\Exp\left[\B{\Sigma}_k(v)|\,v\right]\right)\nonumber\\
	&=-\sum_{i=1}^{d_k} \log_2(\lambda_{k,i})=-\log_2\left(\prod_{i=1}^{d_k}\lambda_{k,i}\right)\nonumber\\
	&\ge-d_k\log_2\left(\frac{\overline{\MMSE}_k}{d_k}\right).
	\label{eq:MMSELLB}
\end{align}
Equation \eqref{eq:MMSELLB} shows that, contrarily to the single stream scenario, a lower bound based on the average MMSE is not tight. Hence, we introduce the per-stream rate target for the $k$-th user and the $i$-th stream, $\varrho_{k,i}$. Therefore, the QoS constraints are satisfied when
\begin{equation}
  \overline{\text{MMSE}}\mbox{}_{k,i}^\text{BC}(v)=\lambda_{k,i}
	\le 2^{-\varrho_{k,i}}
	\label{eq:PSMMSEtargets}
\end{equation}
where $\rho_k=\sum_{i=1}^{d_k}\varrho_{k,i}$ or, equivalently, when $\prod_{i=1}^{d_k}\lambda_{k,i}\leq 2^{-\rho_k}$. 

We now resort to a nested optimization procedure to solve \eqref{eq:BCform}. The outer procedure finds the optimum way to split the target rate $\rho_k$ into the $d_k$ per-stream target rates $\varrho_{k,i}$ in order to minimize the total transmit power, i.e.
\begin{align}
  \min_{\{\bs{\varrho}_k\}_{k=1}^K} P_T(\bs{\varrho})
	\,\,\text{s. t.}\,\,\mathbf{1}^\Tr\bs{\varrho}_k=\rho_k,
	\,\,\text{and}\,\, \bs{\varrho}_k\ge\mathbf{0}\ \forall k,
	\label{eq:outeropt}
\end{align}
with $\bs{\varrho}=[\bs{\varrho}_1^\Tr,\dots,\bs{\varrho}_K^\Tr]^\Tr$, and $\bs{\varrho}_k=[\varrho_{k,1},\dots,\varrho_{k,d_k}]^\Tr$.

The inner optimization determines the minimum transmit power for given per-stream average rate targets $\bs{\varrho}$, that is, $P_T(\bs{\varrho})$ is the solution to the following variational problem   
\begin{align}
\min_{\left\{\B{P}_k(v),\B{F}_k\right\}_{k=1}^K}&\sum_{k=1}^{K}\left\|\B{P}_k(v)\right\|_\text{F}^2\nonumber\\
&\quad\text{s. t.}\quad \overline{\MSE}\mbox{}_{k,i}^{\text{BC}}(v)
	\leq 2^{-\varrho_{k,i}}\,\forall k,i.
\label{eq:MSavMSEform}
\end{align}
Note that this new formulation allows us to treat the streams of each user as virtual users [see \eqref{eq:diagMMSE}]. Thus, we solve \eqref{eq:MSavMSEform} point-wise for each $v$ as done in \cite{GoJoCaCa13ICASSP}. Note, however, that this optimization is more stringent than the original one due to the per-stream restrictions.

\section{Projected Gradient}
\label{sec:projgrad}
Similarly to \cite{JoVoUt10}, the optimization problem \eqref{eq:outeropt} is solved in the dual MAC. To this end we introduce $\B{t}_{k,i}$, $\B{H}_{k}\B{C}_{\bs{\eta}_{k}}^{-\He/2}$, $\B{g}_{k,i}$ and $\B{n}\sim\mathcal{N}_{\mathbb{C}}(\mathbf{0},\mathbf{I}_N)$ as the precoders, the channel, the equalizer and the noise in the dual MAC, respectively. We now define the average transmit power $\xi_{k,i}=\Exp[\|\B{t}_{k,i}\|_2^2|\,v]$, the normalized precoders $\bs{\tau}_{k,i}=\xi_{k,i}^{-1}\B{t}_{k,i}$, and the expectations $\bs{\mu}_{k,i}=\Exp[\B{H}_k\B{C}_{\bs{\eta}_k}^{-\He/2}\bs{\tau}_{k,i}|\,v]$ and $\B{\Theta}_{k,i}=\Exp[\B{H}_k\B{C}_{\bs{\eta}_k}^{-\He/2}\bs{\tau}_{k,i}\bs{\tau}_{k,i}^\He\B{C}_{\bs{\eta}_k}^{-1/2}\B{H}_k^\He|\,v]$. Let us introduce the scalar equalizers $r_{k,i}$ with $\B{g}_{k,i}=r_{k,i}\gt_{k,i}$. Note that the precoders in the dual MAC are functions of the channel whereas the receivers depend on the imperfect CSI $v$. Hence the average MSE read as
\begin{align}
&\overline{\MSE}_{k,i}^\text{MAC}(v)=1-2\Re\left\lbrace r_{k,i}^*\gt_{k,i}^\He\bs{\mu}_{k,i}\sqrt{\xi_{k,i}}\right\rbrace\\
&+\left|r_{k,i}\right|^2\left(\gt_{k,i}^\He\sum\nolimits_{l=1}^{K}\sum\nolimits_{j=1}^{d_l}\xi_{l,j}\B{\Theta}_{l,j}\gt_{k,i}+\left\|\gt_{k,i}\right\|_2^2\right),\nonumber
\end{align}
and the minimum average MSE is
\begin{align}
&\overline{\text{MMSEs}}\mbox{}^{\text{MAC}}_{k,i}(v)=1-\xi_{k,i}\left|\gt_{k,i}^{\He}\bs{\mu}_{k,i}\right|^2y_{k,i}^{-1},
\label{eq:MSMIMOMMSE_scal} 
\end{align}
with the scalar $y_{k,i}=\gt_{k,i}^\He\sum_{l=1}^{K}\sum_{j=1}^{d_l}\xi_{l,j}\B{\Theta}_{l,j}\gt_{k,i}+\|\gt_{k,i}\|_2^2$.

Consequently, we rewrite the optimization problem \eqref{eq:MSavMSEform} as
\begin{align}
  P_T(\bs{\varrho})&=
	\min_{\left\{\bs{\tau}_{k,i},\tilde{\B{g}}_{k,i},\xi_{k,i}\right\}_{k,i}^{K,d_k}}
	\sum\nolimits_{m=1}^K\sum\nolimits_{n=1}^{d_m} \xi_{m,n}\nonumber\\
	&\text{s. t.}\quad \overline{\text{MMSEs}}\mbox{}^{\text{MAC}}_{k,i}(v)\leq 2^{-\varrho_{k,i}}\,\forall k,\forall i.
	\label{eq:subproblem}
\end{align}
Remember that this latter optimization problem can be solved in a way similar to the case of a single stream per user (see \cite{GoJoCaCa13ICASSP}).

We now propose to solve the optimization problem \eqref{eq:outeropt} by means of a gradient-projection algorithm. In such algorithm, the direction of the gradient is followed, but it is projected onto the set of values fulfilling the original per-user restrictions. Indeed, let us define the update rule of the per-stream rate targets as 
\begin{equation}
	\varrho'_{k,i}=\varrho_{k,i}
	-s\,\frac{\partial P_T(\bs{\varrho})}{\partial\varrho_{k,i}},
\label{eq:grad_step}
\end{equation}
with the step size $s>0$. To compute the gradient in \eqref{eq:grad_step} we first calculate the derivative
of $\overline{\text{MMSEs}}\mbox{}^{\text{MAC}}_{k,i}(v)$ in \eqref{eq:MSMIMOMMSE_scal}
with respect to the power allocation elements $\xi_{m,n}$. For the cases: $m=k$, $n=i$, and $m\neq k$ or $n\neq i$, we get
\begin{align}
\frac{\partial\overline{\text{MMSEs}}\mbox{}^{\text{MAC}}_{k,i}(v)}{\partial\xi_{k,i}}=-\frac{\left|\gt_{k,i}^\He\bs{\mu}_{k,i}\right|^2}{y_{k,i}^{2}}&\left(y_{k,i}\right.\nonumber\\
&\left.-\xi_{k,i}\gt_{k,i}^\He\B{\Theta}_{k,i}\gt_{k,i}\right), \nonumber
\end{align}
and
\begin{equation}
\frac{\partial\overline{\text{MMSEs}}\mbox{}^{\text{MAC}}_{k,i}(v)}{\partial\xi_{m,n}}=\frac{\xi_{k,i}\left|\gt_{k,i}^\He\bs{\mu}_{k,i}\right|^2\gt_{m,n}^\He\B{\Theta}_{m,n}\gt_{m,n}}{y_{k,i}^2},\nonumber
\end{equation}
respectively. Taking into account the transmit power $P_T(\bs{\varrho})$ dependency with respect to the per-stream targets, and that the equality $\overline{\text{MMSEs}}\mbox{}^{\text{MAC}}_{k,i}=2^{-\varrho_{k,i}}$ holds in the solution of \eqref{eq:subproblem}, we get that the gradient
\begin{equation}
\frac{\partial\overline{\text{MMSEs}}\mbox{}^{\text{MAC}}_{k,i}(v)}{\partial\varrho_{l,j}}=\sum_{m=1}^K\sum_{n=1}^{d_m}\frac{\partial\overline{\text{MMSEs}}\mbox{}^{\text{MAC}}_{k,i}(v)}{\partial\xi_{m,n}}\frac{\partial\xi_{m,n}}{\partial\varrho_{l,j}},
\label{eq:MSMMSEderiv}
\end{equation}
is equal to $-\ln(2)2^{-\varrho_{k,i}}$ for $k,i=l,j$, and $0$ for $k,i\neq l,j$. Let us now introduce the Jacobian matrix of $\B{f}(\bs{\xi})=[\overline{\text{MMSEs}}\mbox{}^{\text{MAC}}_{1,1}(v),\ldots,\overline{\text{MMSEs}}\mbox{}^{\text{MAC}}_{K,d_K}(v)]^\Tr$ as follows
\begin{equation}
[\B{J}_{\B{f}}(\bs{\xi})]_{a,b}=\frac{\partial\overline{\MMSE}\mbox{}_{k,i}^\text{MAC}(v)}
{\partial\xi_{l,j}},
\end{equation}
where  $a=\sum_{m=1}^{k-1}d_m+i$, and $b=\sum_{m=1}^{l-1}d_m+j$. Similarly, the matrix comprising the partial derivatives of the total average power with respect to the per-stream rate targets is defined as $\B{J}_{\bs{\xi}}({\bs{\varrho}})=\frac{\partial \bs{\xi}}{\partial \bs{\varrho}^{\Tr}}$.
Hence, we rewrite \eqref{eq:MSMMSEderiv} as 
\begin{equation}
\frac{\partial\overline{\text{MMSE}}\mbox{}^{\text{MAC}}_{k,i}}{\partial\varrho_{l,j}}=\left[\B{J}_{\B{f}}\left(\bs{\xi}\right)\B{J}_{\bs{\xi}}\left(\bs{\varrho}\right)\right]_{a,b}
=-\ln(2)\left[\B{W}\right]_{a,b},
\label{eq:MSMMSEderivMatrix}
\end{equation} 
with $\B{W}=\diag(2^{\varrho_{1,1}},\ldots,2^{\varrho_{1,d_1}},\ldots,2^{\varrho_{K,d_K}})$ being the matrix collecting the inverse of the average MMSE targets. Hence, $\B{J}_{\bs{\xi}}({\bs{\varrho}})$, which contains the partial derivatives necessary for the gradient step in \eqref{eq:grad_step}, is obtained by left multiplying times the inverse of $\B{J}_{\B{f}}(\bs{\xi})$ in \eqref{eq:MSMMSEderivMatrix}, that is
\begin{equation}
\B{J}_{\bs{\xi}}(\bs{\varrho})=-\ln(2)\B{J}_{\B{f}}(\bs{\xi})^{-1}\B{W}.
\end{equation}
Therefore, the update for the $k$-th user $i$-th stream reads as
\begin{equation}
\frac{\partial P_T(\bs{\varrho})}{\partial \varrho_{k,i}}
=-\ln(2)\mathbf{1}^{\Tr}\B{J}_{\B{f}}(\bs{\xi})^{-1}\B{W}\B{e}_{\sum_{m=1}^{k-1}d_m+i},
\label{eq:gradient}
\end{equation}
where $\B{e}_i$ is the  canonical vector.

We now prove that $-\B{J}_{\B{f}}(\bs{\xi})$ is a Z-matrix, i.e., the diagonal elements are positive and the off-diagonal ones are negative. Indeed, let us define the diagonal matrix $\B{D}=\diag(\bs{\xi})$. We now observe that the following inequality $\sum\nolimits_{b\neq a}\left|[-\B{J}_{\B{f}}(\bs{\xi})\B{D}]_{a,b}\right|<[-\B{J}_{\B{f}}(\bs{\xi})\B{D}]_{a,a}$ holds for every row $a=\sum_{j=1}^{k-1}d_j+i$, corresponding to the $k$-th user $i$-th stream. Hence, $\B{J}_{\B{f}}(\bs{\xi})\B{D}$ is strictly diagonally dominant and $-\B{J}_{\B{f}}(\bs{\xi})$ is a non-singular M-matrix with positive inverse \cite{BePl}. This result aligns with the intuition that a lower target rate $\varrho_{l,j}$ also leads to a lower transmit power $P_T(\bs{\varrho})$.

It is important to note that after the target update \eqref{eq:grad_step}, the per-stream targets do not fulfill the original constraints $\sum_{i=1}^{d_k}\varrho_{k,i}^\prime=\rho_k$. Therefore, we propose to perform a projection onto the $k$-th user set of feasible target rates by minimizing the following Euclidean distance
\begin{equation}
\min_{\varrho_{k,i}\geq 0}\sum_{i=1}^{d_k}(\varrho_{k,i}-\varrho'_{k,i})^2\quad\text{s. t.}\quad\sum_{i=1}^{d_k}\varrho_{k,i}=\rho_k.
\label{eq:projForm}
\end{equation}

The \emph{Karush-Kuhn-Tucker} (KKT) conditions of \eqref{eq:projForm} lead to the following projection
\begin{align}
\varrho_{k,i}&=\max\left\{\varrho'_{k,i}-\mu_k,0\right\}\nonumber\\
\mu_k&=\frac{1}{d_k}\left(\sum_{i=1}^{d_k}\varrho'_{k,i}-\rho_k\right).
\nonumber
\end{align}
Note that some of the $k$-th user per-stream targets could be switched off (i.e. $\varrho_{k,i}=0$) after the projection. In such a case, the power assigned to such a user is $\xi_{k,i}=0$ and the corresponding gradient is also zero.  That way, the stream will not be switched on again. Such behavior, observed in \cite{GoJoCaCa14SAM}, is avoided by using ``dummy'' filters so that inactive streams do not cause interference but the transmit and receive filters are updated with $\xi_{k,i}=1$. Consequently, ``dummy'' filters do not affect the entries of $\B{J}_{\B{f}}(\bs{\xi})$ for active streams while the entries of the inactive streams are forced to be non-zero.

\label{sec:MSMIMOalgorithm}
\begin{algorithm}[t]
\caption{Power Minimization Algorithm}
\label{alg:pwrmin}
\begin{algorithmic}[1]
\STATE 
$\ell \leftarrow 0$,
\text{random initialization:}\ $\B{P}_k$ and $\varrho_{k,i}^{(0)}$,
$\B{P}_k\leftarrow\B{P}'_k,\forall k$
\STATE Solve \eqref{eq:MSavMSEform}, i.e. $\B{t}_{k,i}^{(0)}$, $\B{g}_{k,i}^{(0)}$,
and $\xi_{k,i}^{(0)}\,\forall k,i$. 
\REPEAT
\STATE $\ell \leftarrow \ell+1$
\STATE $\delta_{k,i}^{(\ell)}\leftarrow\frac{\partial P_T(\bs{\varrho}^{(\ell-1)})}{\partial \varrho_{k,i}^{(\ell-1)}},\,\forall k,i$ [see (\ref{eq:gradient})],
$b_\text{exit}\leftarrow 0$, $s \leftarrow s_0$
\REPEAT
\STATE $\varrho'_{k,i}\leftarrow\varrho_{k,i}^{\left(\ell-1\right)}-s \delta_{k,i}^{(\ell)},\,\forall k,i$. Gradient step \eqref{eq:grad_step}
\STATE $\varrho_{k,i}^{\left(\ell\right)}\leftarrow \max\{\varrho'_{k,i}-\mu_k^{(\ell)},0\}$ $\forall k,i$. Projection 
\STATE Solve \eqref{eq:MSavMSEform}, i.e. $\B{t}_{k,i}^{(\ell)}$, $\B{g}_{k,i}^{(\ell)}$, and $\xi_{k,i}^{(\ell)},\,\forall k,i$ 
\IF{$P_T^{(\ell-1)}-P_T^{(\ell)}> 0$}
\STATE $b_\text{exit}\leftarrow 1$
\ELSE
\STATE $s \leftarrow\frac{s}{2}$. Step size update
\ENDIF
\UNTIL{$b_\text{exit}=1$}
\UNTIL{$\sum_{k=1}^{K}\sum_{i=1}^{d_k}\xi_{k,i}^{(\ell-1)}-\sum_{k=1}^{K}\sum_{i=1}^{d_k}\xi_{k,i}^{(\ell)}\leq \gamma$} 
\end{algorithmic}
\end{algorithm}

\subsection{Proposed Algorithm}
Algorithm \ref{alg:pwrmin} implements the solution described previously for the power minimization in the multiple stream MIMO BC. In line 1 both the precoders and the per-stream rate targets are randomly initialized.  The power minimization \eqref{eq:MSavMSEform} is solved via the methods proposed in \cite{GoJoCaCa13ICASSP,GoJoCaCA13_2} since every stream is treated as a virtual user (see line 2). The algorithm performs a steepest descent method, for which the gradient is computed in line 5. Line 7 updates the per-stream target rates $\varrho_{k,i}$ according to (\ref{eq:grad_step}) and the projection to the set of feasible solutions is implemented in line 8 (see the proposed solution for \eqref{eq:projForm}).

Next, the power minimization \eqref{eq:MSavMSEform} is updated (see line 9). Then, if the BC total power is smaller than that achieved in the previous iteration, the per-stream target rates and the corresponding transmit and receive filters are updated. If not, the step size $s$ is reduced in line 13. If the initial QoS constraints are feasible, the convergence to a local minimum is guaranteed since in every iteration the power decreases or remains unchanged. Finally, we set the threshold $\gamma$ in line 16 to check whether convergence is reached or not.

\begin{figure}[!t]
  \centering
	\includegraphics[width=\linewidth]{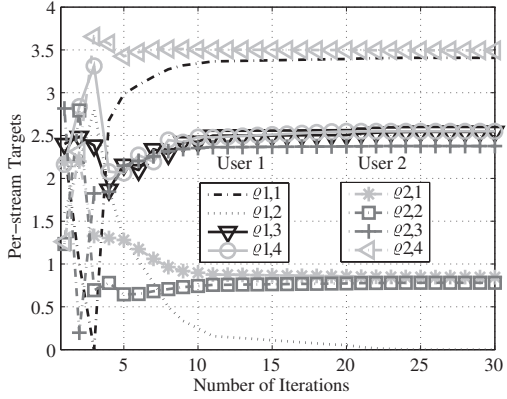}
	\caption{ Average Sum Power (dB) vs. Number of Iterations.}
\label{fig:targets}
\end{figure}

For feasibility testing we generalize the single-stream vector channel  procedure employed in \cite{GoJoCaCA13_2} to the multiple-stream MIMO channel. We thereby obtain the matrix 
\begin{equation}
\B{E}=\mathbf{I}_{d}-\Exp\left[\B{\Upsilon}^\He|v\right]\left(\Exp\left[\B{\Upsilon}\B{\Upsilon}^\He|v\right]+\sigma^2\mathbf{I}_N\right)^{-1}\Exp\left[\B{\Upsilon}|v\right],
\label{eq:MMSEsmatrix}
\end{equation} 
using $\B{T}_l=[\B{t}_{l,1},\ldots,\B{t}_{l,d_l}]$, $\B{\Upsilon}_l=\B{H}_l\B{T}_l\in\mathbb{C}^{N\times d_l}$ and $\B{\Upsilon}=[\B{\Upsilon}_1,\ldots,\B{\Upsilon}_K]\in\mathbb{C}^{N\times d}$, with the total number of data streams $d=\sum_{k=1}^{K}d_k$. Note that $\trace(\B{E})$ is the average sum-MMSE, and $\B{E}$ contains the average MMSEs for the $k$-th user $i$-th stream, $\overline{\MMSE}_{k,i}$, in the entry $[\B{E}]_{a,a}$, with $a=\sum_{l=1}^{k-1}d_l+i$. Accordingly, the $k$-th user average MMSE, $\overline{\MMSE}_k$, corresponds to $\trace([\B{E}]_{b:c,b:c})$, with $b=\sum_{l=1}^{k-1}d_l+1$ and $c=\sum_{l=1}^{k}d_l$. When setting $\sigma^2=0$, $\trace(\B{E})$ gives the  sum-MMSE lower bound for the set of precoders $\{\B{T}_k\}_{k=1}^K$, i.e., feasible MMSE targets have to fulfill 
\begin{align}
\sum_{k=1}^{K}\sum_{i=1}^{d_k}&2^{-\varrho_{k,i}}\leq d\label{eq:feasib}\\ &-\trace(\Exp\left[\B{\Upsilon}^\He|v\right]\left(\Exp\left[\B{\Upsilon}\B{\Upsilon}^\He|v\right]\right)^{-1}\Exp\left[\B{\Upsilon}|v\right]).\nonumber 
\end{align}
Note that if $\bs{\varrho}$ are feasible, any distribution between the streams $\bs{\varrho}^\prime$ such that $\sum_{k=1}^{K}\sum_{i=1}^{d_k}2^{-\varrho_{k,i}^\prime}=\sum_{k=1}^{K}\sum_{i=1}^{d_k}2^{-\varrho_{k,i}}$ satisfies the inequality \eqref{eq:feasib}.

\section{Simulation Results}
\label{sec:sim}

To illustrate the performance of Algorithm \ref{alg:pwrmin} we have considered a MIMO BC with $K=2$ users, $R=6$ receive antennas per user and $N=8$ transmit antennas. Each user allocates $d_1=d_2=4$ streams. The AWGN is zero-mean with $\B{C}_{\bs{\eta}}=\mathbf{I}_R$, and the per-user target rates are set to $\rho_1=8.5$ and $\rho_2=7.5$ bits per channel use. The step size is $s_0=2$, and the stop threshold is fixed to $\gamma=10^{-5}$. 
We assume the following error model 
 \begin{equation}
\B{H}_k=\bar{\B{H}}_k+\tilde{\B{H}}_k,
\label{eq:MIMOchannel_model}
\end{equation}
with $\bar{\B{H}}_k=\Exp[\B{H}_k|\,v]$ and $\tilde{\B{H}}_k$  being the imperfect CSI error, with $\tilde{\B{H}}_k\sim\mathcal{N}_\mathbb{C}(\mathbf{0},\B{C}_{\tilde{\B{H}}_k})$, where $\B{C}_{\tilde{\B{H}}_k}=\Exp[(\B{H}_k-\bar{\B{H}}_k)(\B{H}_k-\bar{\B{H}}_k)^\He|\,v]$. We consider first and second order moments $[\Exp[\B{H}_k|v]]_{1:N,r}=\B{u}_{k,r}$, for each $r\in\{1,\ldots,R\}$ with $u_{k,r,n}=\eul^{\imj(n-1)\varphi_k}$ and $\varphi_k\sim \mathcal{U}(0,2\pi)$, and $\B{C}_{\tilde{\B{H}}_k}=R\mathbf{I}_{N},\,\forall k$. Recall that no closed form expressions for the expectations in \eqref{eq:MSMIMOMMSE_scal} have been found. Therefore, we employ Monte Carlo numerical integration with $M=1000$ channel realizations. This way, we calculate
\begin{align}
&\bs{\mu}_{k,i}=\frac{1}{M}\sum_{m=1}^{M}\B{H}_k^{(m)}\B{C}_{\bs{\eta}_k}\bs{\tau}_{k,i}^{(m)}\nonumber\\
&\B{\Theta}_{k,i}=\frac{1}{M}\sum_{m=1}^{M}\B{H}_k^{(m)}\B{C}_{\bs{\eta}_k}^{-\He/2}\bs{\tau}_{k,i}^{(m)}\bs{\tau}_{k,i}^{(m),\He}\B{C}_{\bs{\eta}_k}^{-\frac{1}{2}}\B{H}_k^{(m),\He}\nonumber
\end{align} 
 where $\bs{\tau}_{k,i}^{(m)}$ is the normalized MAC precoder for the channel realization $m$.

\begin{figure}[!t]	
	\includegraphics[width=\linewidth]{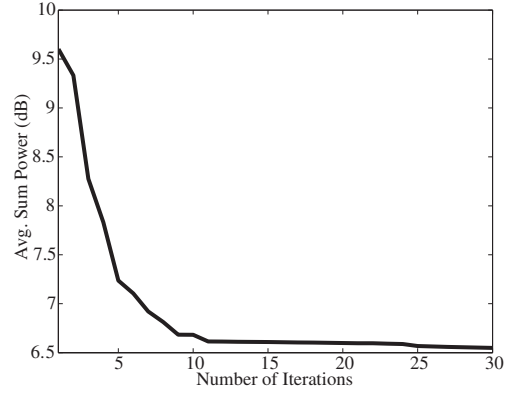}
\caption{ Per-Stream Targets vs. Number of Iterations.}
\label{fig:power}
\end{figure}

The evolution of the per-stream target rates can be observed in  Fig. \ref{fig:targets}. The proposed method makes it possible for the stream $1$ of user $1$ to be deactivated and afterwards activated (see iterations $3$ and $4$). Note that at each iteration the sum of the per-stream targets for user $k$ is equal to $\rho_k$.
The total power needed to achieve those targets is shown in  Fig. \ref{fig:power}. It can be seen from the figure how the power is gradually reduced. Observe that both the per-stream targets and the total transmit power converge at about $15$ iterations and, after iteration $9$, the power reduction is negligible. 

\section{Conclusion}
\label{sec:conc}
This work addresses the power minimization of the multiple-stream MIMO BC subject to per-user average rate restrictions. Moreover, the practical assumption of imperfect CSI at the transmitter is considered, leading to a complicated problem formulation. To tackle with this difficulty, we propose to reformulate the problem by introducing average MMSE-based conservative restrictions. Moreover, the target for each user is distributed between the streams. To find the less power consuming distribution, a projected gradient method is proposed. By using this procedure, convergence to a locally optimum solution is ensured. The flexibility of the proposed algorithm is improved by allowing the streams to switch on and off for convenience.


\acks This work was supported by Xunta de Galicia, MINECO of Spain,
and FEDER funds of the EU under grants 2012/287 and  TEC2013-47141-C4-1-R.

%
%
%

\bibliographystyle{wileyj}
\bibliography{ettdoc}

\biogs 
\textbf{Jos\'e P. Gonz\'alez Coma}
 was born in Mar\'in, Spain, in 1986. He received the Engineering Computer, M.Sc., and PhD. degrees in 2009, 2010 and 2015 from Universidade da Coru\~na, Spain. Since 2009 he is with the Grupo de Tecnolog\'ia Electr\'onica y Comunicaciones (GTEC) at the Departamento de Electr\'onica y Sistemas. Currently, he is awarded with a FPI grant from Ministerio de Ciencia e Innovaci\'on. His main research interests are in designs of limited feedback and robust precoding techniques in MIMO wireless communication systems. Since Sep 2013 he is assistant professor in Universidade da Coru\~na.
 \bigbreak
 
 \textbf{Michael Joham}
 was born in Kufstein, Austria, 1974. He received the Dipl.-Ing.
 and Dr.-Ing. degrees (both summa cum laude) in electrical engineering
 from the Technische Universit\"{a}t M\"{u}nchen (TUM), Germany,
 in 1999 and 2004, respectively.
 
 Dr. Joham was with the Institute for Circuit Theory and Signal Processing
 at the TUM from 1999 to 2004. Since 2004, he has been with the Associate
 Institute for Signal Processing at the TUM, where he is currently a senior
 researcher. In the summers of 1998 and 2000, he visited the Purdue University,
 IN. In spring 2007, he was a guest lecturer at the University of A coru\~na,
 Spain. In spring 2008, he was a guest lecturer at the University of the German
 Federal Armed Forces in Munich, Germany, and a guest professor at the
 University of A coru\~na, Spain. In Winter 2009, he was a guest lecturer
 at the University of Hanover, Germany. In Fall 2010 and 2011,
 he was a guest lecturer at the Singapore Institute of Technology.
 His current research interests are
 precoding in mobile and satellite communications, limited rate feedback,
 MIMO communications, and robust signal processing.
 
 Dr. Joham received the VDE Preis for his diploma thesis in 1999 and the
 Texas-Instruments-Preis for his dissertation in 2004. In 2007, he was a
 co-recipient of the best paper award at the International ITG/IEEE Workshop
 on Smart Antennas in Vienna. In 2011, he received the ITG Award 2011
 of the German Information Technology Society (ITG).
 \bigbreak
 
 \textbf{Paula M. Castro}
  was born in Orense, Spain, 1975. She received her MS and PhD degrees in Electrical Engineering (2001) from the University of Vigo, Spain, and in Computer Engineering (2009) from the University of A Coru\~na, Spain, respectively. Since 2002 she is with the Department of Electronics and Systems at the University of A Coru\~na where she is currently an Associate Professor. Paula Castro is coauthor of more than forty papers in peer-reviewed international journals and conferences. She has also participated as a research member in more than thirty five research projects and contracts with the regional and national governments. She is a co-recipient of the Best Paper Award at the International ITG/IEEE Workshop on Smart Antennas, Vienna, 2007. Her research is currently devoted to the design of multiuser wireless communications systems.
  
   \bigbreak
  \textbf{Luis Castedo}
  was born in Santiago de Compostela, Spain, in 1966.
  He received the Ingeniero de Telecomunicaci\'on and Doctor
  Ingeniero de Telecomunicaci\'on degrees, both from Universidad
  Polit\'ecnica de Madrid (UPM), Spain, in 1990 and 1993, respectively.
  Between November 1994 and October 2001, he has been an
  Associate professor at the Departamento de Electr\'onica y Sistemas
  at Universidad de A Coru\~na, Spain, where he is currently Full
  Professor. He has been chairman of the Department between 2003
  and 2009.
  
  Luis Castedo is coauthor of more than one hundred and fifty papers
  in peer-reviewed international journals and conferences. He has
  also been principal researcher in more than thirty research projects
  funded by public organisms and private companies. His research interests are signal processing and digital
  communications with special emphasis on blind adaptive filtering,
  estimation/equalization of MIMO channels, space-time coding and
  prototyping of digital communication equipments.
\end{document}